\documentclass[aps,prd,nofootinbib,amsmath,superscriptaddress]{revtex4-2}

\usepackage{graphicx}
\usepackage{dcolumn}
\usepackage{bm}
\usepackage{epsfig}
\usepackage{amsmath}
\input{epsf}
\usepackage{psfrag}
\usepackage{xcolor}
\usepackage{slashed}
\usepackage{ulem}
\usepackage{CJK}
\usepackage{subcaption}
\usepackage{ragged2e}
\newcommand{\bea}{\begin{eqnarray}}
\newcommand{\eea}{\end{eqnarray}}
\newcommand{\be}{\begin{equation}}
\newcommand{\ee}{\end{equation}}
\newcommand{\nn}{\nonumber}

\usepackage[compatibility=false]{caption}
\begin{document}
\begin{CJK*}{UTF8}{gbsn}
\title{Enhanced tripartite entanglement near radiation zeroes}

\author{Haotian Cao (曹浩天)}

\email{haotiao.cao@northwestern.edu}
 \affiliation{Department of Physics \& Astronomy, Northwestern University, Evanston, Illinois 60208, USA}
\affiliation{Center for Frontiers in Nuclear Science, Stony Brook University, Stony Brook, NY 11794, USA}
\author{Frank~Petriello}
\email{f-petriello@northwestern.edu}
 \affiliation{Department of Physics \& Astronomy, Northwestern University, Evanston, Illinois 60208, USA}

\begin{abstract}
We study tripartite entanglement among the two $\tau$ spins and the photon polarization in $e^+e^-\to \tau\bar{\tau}\gamma$. The scattering amplitude in the soft-photon approximation vanishes at specific kinematic configurations known as radiation zeros. We show that the next-to-leading soft term also vanishes for opposite $\tau$ helicities, leaving mass-dependent amplitudes for equal $\tau$ helicities that lead to a strong enhancement of tripartite entanglement. At the radiation zero where the photon is moving nearly perpendicular to the beam, two complementary helicity amplitudes become nearly equal in magnitude, yielding a GHZ-like state.
In the full tree-level calculation at Belle-II kinematics, we find $F_3\simeq0.9985$ near this radiation zero for fixed incoming helicities.
Although the differential cross section decreases near the radiation zero, the significant integrated luminosity at Belle-II makes this effect potentially observable. 

\end{abstract}

\maketitle

\end{CJK*}

\section{Introduction}

Entanglement is a critical aspect of quantum information science (QIS). It is a key feature that helps enable computations not possible with classical computers and generates much of the excitement for technological uses of quantum computation in the future. The importance of entanglement for future QIS applications has motivated its study in areas not traditionally formulated in this way, such as in collider experiments. For example, spin entanglement in $t\bar{t}$ was recently observed at the Large Hadron Collider (LHC)~\cite{ATLAS:2023fsd,CMS:2024pts,CMS:2024zkc}. This is just one result illustrating the growing interest in QIS within the high-energy physics community. Over the past few years, the application of QIS in high energy theory has offered a promising new perspective for studying fundamental interactions and searching for physics beyond the Standard Model (SM)~\cite{Duch:2024pwm,Fabbrichesi:2023jep, Bernal:2023ruk, Fabbrichesi:2024xtq, Fabbrichesi:2024wcd, Fabbrichesi:2025ywl,Fuks:2021xje,Maltoni:2024tul,Fuks:2024yjj,Cao:2025xnp}. High-energy collisions also provide a way to probe quantum correlations at energy scales and through interactions that are difficult to access in conventional quantum information experiments~\cite{Afik:2020onf,Severi:2021cnj,Barr:2024djo}. 

Most studies of entanglement have focused on bipartite entanglement, the entanglement of two qubits. Bipartite entanglement and related quantum observables have been widely studied at colliders, with extensive work on top-quark pairs~\cite{Afik:2020onf,Afik:2022kwm,Altomonte:2024upf,Durupt:2025wuk,Fabbrichesi:2021npl,Severi:2021cnj,Aguilar-Saavedra:2022uye,Han:2023fci,Dong:2023xiw,Cheng:2023qmz,Cheng:2024btk,Fabbrichesi:2025psr,Maltoni:2024csn,Choi:2026omc,Low:2025aqq,Nason:2025hix,Dong:2024xsg,Dong:2024xsb,Afik:2022dgh,Han:2024ugl,White:2024nuc,Aguilar-Saavedra:2023hss,Aoude:2022imd,Severi:2022qjy,Maltoni:2024tul,ATLAS:2023fsd,CMS:2024pts,CMS:2024zkc,Han:2026woa}, $\tau\bar{\tau}$ pairs~\cite{Ehataht:2023zzt,Han:2025ewp,Fabbrichesi:2024wcd,Zhang:2025mmm,Fabbrichesi:2022ovb,Fabbrichesi:2025ywl,Altakach:2022ywa,Ma:2023yvd,Jeans:2026eys,Lee:2026wlk}, and pairs of weak gauge bosons produced in scattering processes or in Higgs decays~\cite{Fabbrichesi:2023cev,Morales:2023gow,Bi:2023uop,Grossi:2024jae,Goncalves:2025mvl,Ruzi:2025jql,Ding:2025mzj,Aoude:2023hxv,Pelliccioli:2026ltl,Barr:2021zcp,Aguilar-Saavedra:2022wam,Aguilar-Saavedra:2022mpg,Ashby-Pickering:2022umy,Fabbri:2023ncz,Wu:2024ovc,Ruzi:2024cbt,Bernal:2024xhm,De:2025dpo,Bernal:2023ruk,Fabbrichesi:2023jep,Sullivan:2024wzl,DelGratta:2025qyp,Aguilar-Saavedra:2025byk,Goncalves:2025xer,Goncalves:2026njf,Aguilar-Saavedra:2025njw,Aguilar-Saavedra:2024jkj,CMS-PAS-HIG-25-011,ATLAS:2026hye}. Bipartite quantum correlations have also been studied in a variety of other settings~\cite{Afik:2025grr,Cheng:2025cuv,Lin:2025eci,Hatta:2026dqs,Qi:2025onf,Fucilla:2025kit,Fucilla:2026mkg,Hatta:2025obw,Cheng:2025zaw,Bloss:2026yrf,Xiao:2026tbs,Altakach:2026fpl,Guo:2026yhz,Fang:2026ddi,Aoude:2025ovu,Gu:2025ijz,Aoude:2026eeg,Agrawal:2026zwa,Fucilla:2026zhy}. The next step beyond this involves tripartite entanglement of three qubits. Tripartite systems with three qubits play an important role in QIS, providing the simplest setting in which genuine multipartite entanglement can arise and serve as a resource for quantum information protocols~\cite{Cunha:2019jex,Walter:2016lgl}. However, they have received far less attention than bipartite systems in high-energy physics. 
Early studies examined tripartite entanglement in both particle decays and scattering processes~\cite{Sakurai:2023nsc}, including ortho-positronium decays~\cite{Acin:2000cs,Hiesmayr:2017xgx}, neutral kaon decay~\cite{Fabbrichesi:2025zpw}, Higgs boson decays~\cite{Aguilar-Saavedra:2024whi,Morales:2024jhj,Banacki:2026msu}, $e^+e^-\rightarrow t\bar{t}Z$~\cite{Goncalves:2026obl}, Bethe-Heitler process~\cite{Cao:2026zqo} and several other $2\to 3$ and $3\to 3$ processes~\cite{Blasone:2024dud,Sou:2025tyf,Chu:2026yxm,Cao:2026mza}. Some general features of these studies show that tripartite entanglement can occur for certain vector-like and tensor-like four-fermion couplings~\cite{Sakurai:2023nsc}, and that there may be limited sensitivity to GHZ-like tripartite entanglement in future high-energy electron-positron colliders~\cite{Goncalves:2026obl}. The spin state in $H\to\ell^+\ell^-\gamma$ takes the GHZ form when the photon is collinear with either lepton~\cite{Morales:2024jhj}. Calculations of the Bethe-Heitler amplitude yield both GHZ and W states~\cite{Cao:2026zqo}. The generation of GHZ states in processes accessible at existing colliders remains relatively unexplored.

Our goal in this paper is to point out that tripartite entanglement can be enhanced at special kinematic points known as radiation zeroes. Radiation zeros were first identified in $W\gamma$ production, where the tree-level amplitude vanishes at a specific scattering angle~\cite{Mikaelian:1979nr,Brown:1979ux}. Subsequent work explained this zero and derived a general cancellation condition for tree-level single photon emission in terms of the charges and momenta of the external particles~\cite{Goebel:1980es,Brodsky:1982sh,Brown:1982xx}. Zeros that satisfy this condition are now known as type-I zeros and require the external charged particles to carry charges of the same sign. A second class of radiation zero was found in lepton-quark scattering and $e^+e^-$ annihilation into quark pairs with single photon emission~\cite{Heyssler:1997fy,Heyssler:1997ng}. These zeros can occur in processes involving particles of opposite charge. They arise from cancellations between radiation amplitudes at specific kinematic configurations in which all external three-momenta lie in the same plane. These are known as type-II radiation zeros, and are the focus of this work.

We explain here the basic physical mechanism that makes radiation zeroes interesting from the perspective of tripartite entanglement. We illustrate this for $e^+e^-\rightarrow \tau \bar{\tau}\gamma$ at Belle-II center-of-mass energy, for which there is a large amount of existing data. At fixed momenta, the leading soft-photon amplitude factorizes, leaving the photon polarization unentangled with the spins of the $\tau$ pair. This contribution vanishes at the type-II radiation zero. The residual depends on the finite $\tau$-mass. These surviving contributions come from the equal-helicity $\tau\bar{\tau}$ channels. For fixed incoming helicities, at a zero where the photon is emitted nearly perpendicular to the beam, the two dominant helicity amplitudes, $+_{\tau}+_{\bar{\tau}}-_\gamma$ and $-_{\tau}-_{\bar{\tau}}+_\gamma$, become nearly equal in magnitude. Their coherent superposition approaches a GHZ-like state with genuine tripartite entanglement. Although the differential cross section decreases at this point the significant integrated luminosity at Belle-II makes this effect observable. 

Our goal in this paper is to demonstrate the appearance of GHZ-like tripartite entanglement near radiation zeroes in the $e^+e^-\rightarrow \tau \bar{\tau}\gamma$ process. We analytically derive the argument outlined above, and present numerical evidence for $F_3$ near its maximum value of one. We study the observability of this effect given realistic Belle-II luminosities. Although the cross section dips at the radiation zero, GHZ-like entanglement persists away enough from that point. Our estimates of the event yields in the surrounding region for the Belle-II integrated luminosity suggest that this entanglement may be observable. Our paper is organized as follows. We briefly review three-qubit entanglement in Sec.~\ref{sec:QIS}. In Sec.~\ref{sec:rz}, we review radiation zeros and discuss how entangled final states arise at these zeros. We present our numerical results in Sec.~\ref{sec:num} and summarize our findings in Sec.~\ref{sec:con}.

\section{Review of three-qubit entanglement }
\label{sec:QIS}

We begin by reviewing the formalism for three-qubit entanglement relevant for the process $e^+e^- \to \tau\bar{\tau}\gamma$. We denote the helicities of the final-state particles as  $h_{\tau}$, $h_{\bar{\tau}}$, and $h_\gamma$, respectively. Unless stated otherwise we work in a basis of longitudinal polarizations for the final-state $\tau$ leptons and circular polarizations for the photon. When we fix the external momenta and the incoming helicities, we can write the final-state spin state in Hilbert space as
\bea
|\phi\rangle=\sum_{h_{\tau}h_{\bar{\tau}}h_\gamma} M_{\lambda_{e^-}\lambda_{e^+}}^{h_{\tau}h_{\bar{\tau}}h_\gamma}|h_{\tau}h_{\bar{\tau}}h_\gamma\rangle
\eea
The $M$ in this expression denotes the standard matrix element computed in perturbation theory. We focus here on tree-level results in QED, since we will be interested in collisions far below the $Z$-pole. The corresponding unnormalized $8\times8$ density matrix is
\bea
R_{\lambda_{e^-}\lambda_{e^+}}^{h_{\tau}h_{\bar{\tau}}h_\gamma,h_{\tau}'h_{\bar{\tau}}'h_\gamma'}\equiv
M_{\lambda_{e^-}\lambda_{e^+}}^{h_{\tau}h_{\bar{\tau}}h_\gamma}\left[M_{\lambda_{e^-}\lambda_{e^+}}^{h_{\tau}'h_{\bar{\tau}}'h_\gamma'}\right]^*.
\eea
We have denoted the initial-state helicities as $\lambda_{e^-}$ and $\lambda_{e^+}$. The normalized density matrix can then be written as 
\be
\rho_{\lambda_{e^-}\lambda_{e^+}}\equiv R_{\lambda_{e^-}\lambda_{e^+}}/\text{Tr}\left[R_{\lambda_{e^-}\lambda_{e^+}}\right]
\ee

For a three-qubit state, stochastic local operations and classical communication (SLOCC) give rise to six equivalence classes of pure states~\cite{Dur:2000zz}. There is one fully separable class, three biseparable classes, one GHZ class~\cite{Greenberger:1989tfe}, and one W class~\cite{Dur:2000zz}.
The fully separable states can be written as $|\phi_A\rangle \otimes |\phi_B\rangle \otimes |\phi_C\rangle$; they are not entangled. 
A biseparable three-qubit pure state can be written as the tensor product of an entangled two-qubit pure state and a pure state of the remaining qubit. An example of such a state is$
\frac{|-_A+_B\rangle-|+_A-_B\rangle }{\sqrt{2}}\otimes |\psi_C\rangle$. Such states do not represent genuine tripartite entanglement.

To proceed further we quantify the entanglement between two states $A$ and $B$ by the concurrence of their reduced state $\rho_{AB}=\operatorname{Tr}_C\rho$. The concurrence is~\cite{Wootters:1997id}
\be
\mathcal C_{AB}=\max\{0,r_1-r_2-r_3-r_4\}\, ,
\ee 
where $r_1\ge r_2\ge r_3\ge r_4$ are the square roots of the eigenvalues of $\rho_{AB}(\sigma_y\otimes\sigma_y)\rho_{AB}^{*}(\sigma_y\otimes\sigma_y)$. 
The entangled two-qubit pure states are characterized by the non-zero concurrence ${\mathcal C}_{AB}\neq0$~\cite{Wootters:1997id}.
The entanglement between one qubit and the rest of the system can be characterized by the concurrence ${\mathcal C}_{A(BC)}=\sqrt{2(1-{\rm Tr}\rho_A^2)}$, where $\rho_A=\text{Tr}_{BC}[\rho]$~\cite{Rungta:2001zcj}. The negativity $N(A| BC)$ can also quantify the entanglement between qubit $A$ and the joint subsystem $BC$. It is defined as $N(A| BC)=(\|\rho^{T_A}\|_1-1)/2$~\cite{Vidal:2002zz}, where $T_A$ denotes partial transposition with respect to $A$ and $\|\cdot\|_1$ is the trace norm.
For a pure three-qubit state, $N(A| BC)=\mathcal C_{A(BC)}/2$.
For a pure three-qubit state, the monogamy bound is ${\mathcal C}_{AB}^2+{\mathcal C}^2_{AC} \le {\mathcal C}^2_{A(BC)} \le 1$~\cite{Coffman:1999jd}. This means that, if two qubits are maximally entangled, neither of them can be simultaneously entangled with the other one. 

These statements guide us toward an appropriate definition of tri-partite entanglement.  The difference $\tau_3={\mathcal C}^2_{A(BC)}-{\mathcal C}_{AB}^2-{\mathcal C}^2_{AC}$ is called the three-tangle~\cite{Coffman:1999jd}. A pure three-qubit state belongs to the GHZ class if and only if the three-tangle is non zero~\cite{Dur:2000zz}. A representative of this class is the GHZ state: $\frac{|+++\rangle- |---\rangle}{\sqrt{2}}$. In the GHZ state, each qubit is maximally entangled with the other two ${\mathcal C}_{A(BC)}={\mathcal C}_{B(CA)}={\mathcal C}_{C(AB)}=1$. From the monogamy bound we know that there will not be two qubits that are pairwise entangled: ${\mathcal C}_{AB}={\mathcal C}_{BC}={\mathcal C}_{CA}=0$. We quantify genuine tripartite entanglement using the normalized area of the concurrence triangle $F_3$~\cite{Jin:2022kxb}:
\bea
 F_3=\left[\frac{16}{3}Q(Q-{\mathcal C}_{A(BC)})(Q-{\mathcal C}_{B(CA)})(Q-{\mathcal C}_{C(AB)})\right]^{\frac{1}{2}}\, ,
\eea
where $Q=\frac{1}{2}({\mathcal C}_{A(BC)}+{\mathcal C}_{B(CA)}+{\mathcal  C}_{C(AB)})$. The GHZ state exhibits genuine tripartite entanglement, with  $F_3^{\rm GHZ}=1$. For a mixed three-qubit state $\rho$, the convex-roof extension is defined as~\cite{Jin:2022kxb} $F_3=\min\sum_i p_i F_3(|\psi_i\rangle)$ where the minimum is taken over all pure-state decompositions $\rho=\sum_i p_i|\psi_i\rangle\langle\psi_i|$.

When the three-tangle $\tau_3=0$ but ${\mathcal C}_{A(BC)}>0$, ${\mathcal C}_{B(AC)}>0$, and ${\mathcal C}_{C(AB)}>0$, the state is in the W class~\cite{Dur:2000zz}. Pure states in the $W$ class also have genuine tripartite entanglement. A representative of this class is the W state: $\frac{|++-\rangle+ |+-+\rangle+ |-++\rangle}{\sqrt{3}}$. For this state $F_3^{\rm W}=8/9$. 

\section{Radiation zeroes in scattering processes}
\label{sec:rz}

Radiation zeroes occur when the emission of additional photons in a scattering process leads to the vanishing of the differential cross section. They were originally found in studies of $W$-boson production in association with a photon in partonic collisions, when considering anomalous magnetic moments of the $W$-boson~\cite{Mikaelian:1979nr,Brown:1979ux}. Their general explanation in the context of interference effects and the eikonal limit in gauge theories was given later~\cite{Goebel:1980es,Brodsky:1982sh,Brown:1982xx}. Two types of radiation zeroes have been identified in the literature. The traditional radiation zero is known as type-I. It can be derived directly from the soft-photon approximation and occurs only for like-charge scattering, which is not the case here. Type-II zeroes were discovered later~\cite{Heyssler:1997fy,Heyssler:1997ng} and can occur for the opposite-sign scattering relevant here.

To proceed we define our scattering process as $e^-(p_1)e^+(p_2)\rightarrow \tau(p_3)\bar{\tau}(p_4)\gamma(q)$ in the center of mass frame of the colliding particles. The signs label the incoming helicities with respect to their direction. We consider this particular helicity combination as an example, the discussion is similar for the other non-vanishing amplitudes.
We set $m_e=0$ and define $s=(p_1+p_2)^2$, $\omega=q^0$, and $\hat q=q/\omega$. In the soft-photon limit, the matrix elements for one soft-photon emission can be written through next-to-leading power as
\be
\label{eq:exact-rearrangement}
  M=e^3\left(\frac{1}{\omega}M_{\text{LP}}+M_{\text{NLP}}+O\left(\frac{\omega}{s}\right)\right)\,,
\ee
where we defined $\omega$ as the photon energy and
\bea
\label{eq:M_LP}
 M_{\text{LP}}^{h_{\tau}h_{\bar{\tau}}h_\gamma}&\equiv&M_0 ^{h_{\tau}h_{\bar{\tau}}}\mathcal{S}^\mu_{\text{LP}}\varepsilon^*_\mu(q) \, ,\nn
\\
{\mathcal{S}}^\nu_{\text{LP}}&\equiv&\frac{p_3^\nu}{p_3\cdot \hat{q}}-\frac{p_4^\nu}{p_4\cdot \hat{q}}-\frac{p_1^\nu}{p_1\cdot \hat{q}}+\frac{p_2^\nu}{p_2\cdot \hat{q}} \, ,
\nn\\
M_0^{h_{\tau}h_{\bar{\tau}}}&\equiv&\frac{1}{s} \bigl[\bar v_e(p_2)\gamma^\mu u_e(p_1)\bigr]
 \bigl[\bar u_\tau(p_3)
 \gamma_\mu v_\tau(p_4)\bigr]\, ,
 \nn \\
  M_{\text{NLP}}^{h_{\tau}h_{\bar{\tau}}h_\gamma}
&\equiv&
\frac{
 \bigl[\bar v_e(p_2)\gamma^\mu u_e(p_1)\bigr]
}{
 s
}
\bar u_\tau(p_{3})\left[\frac{
 \slashed{\varepsilon}^{*}(q)
 \slashed{q}
 \gamma_\mu 
}{
 2p_{3}\cdot q
}
-\frac{
\gamma_\mu
 \slashed{q}
 \slashed{\varepsilon}^{*}(q)
}{
 2p_{4}\cdot q
}\right] v_\tau(p_{4})
\nn\\
&+&\frac{
 \bigl[\bar u_\tau(p_{3})\gamma_\mu v_\tau(p_{4})\bigr]
}{
  s
}\bar v_e(p_2)\left[\frac{
 \gamma^\mu
 \slashed{q}
 \slashed{\varepsilon}^{*}(q)
}{
  2 p_1\cdot q
}
-\frac{
 \slashed{\varepsilon}^{*}(q)
 \slashed{q}
 \gamma^\mu 
}{
  2p_2\cdot q
}\right]u_e(p_1)
\nn \\ 
&+&\frac{2}{\sqrt{s}}M_0^{h_{\tau}h_{\bar{\tau}}}\left(-\frac{p_1^\nu}{p_1\cdot \hat{q}}+\frac{p_2^\nu}{p_2\cdot \hat{q}}\right)\varepsilon^*_\nu(q)
\,.
\eea
We have used the Dirac equation to rearrange the NLP expression. Although we will not need it here, we note that the leading-power and next-to-leading power results can be expressed in a process-independent representation~\cite{Low:1958sn,Burnett:1967km}.

\begin{figure}[t]
    \centering
    \includegraphics[width=0.9\linewidth]{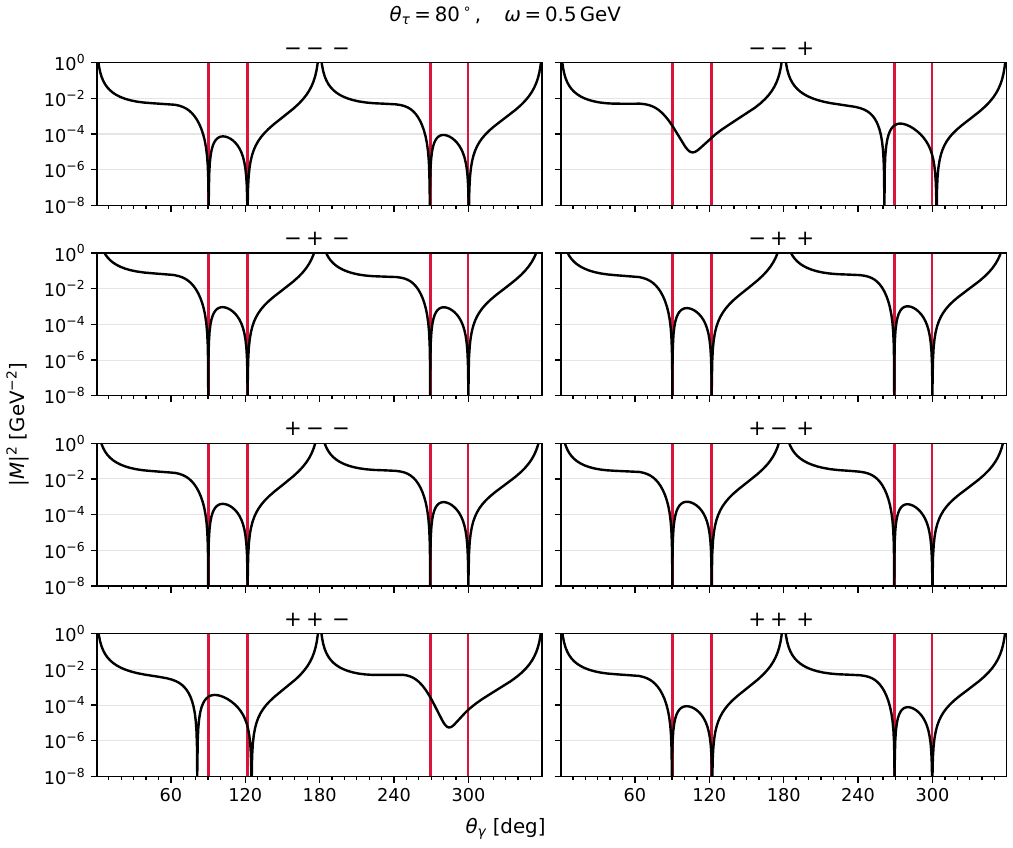}
    \caption{Helicity-resolved matrix element square at $\sqrt{s}=10.58\,\mathrm{GeV}$, $\theta_\tau=80^\circ$, and $\omega=0.5\,\mathrm{GeV}$. The red lines mark the radiation zeros. The final-state helicity labels follow the order $(\tau,\bar{\tau},\gamma)$.}
    \label{fg:2zeros}
\end{figure}

We begin by solving for the location of the radiation zero for both photon helicities at leading power: 
\bea
\label{eq:lprz}
\mathcal{S}^\mu_{\text{LP}}\varepsilon^*_\mu(q) =0\,.
\eea
 This gives type-II radiation zeros~\cite{Heyssler:1997fy}, which are characterized by all momenta lying in the same plane. We assume all momenta are in the $xz$ plane and parameterize them as
\bea
p_1 &=& \frac{\sqrt{s}}{2}(1,0,0,1), \quad p_2 = \frac{\sqrt{s}}{2}(1,0,0,-1),\quad  p_3 =E_3(1,\beta_\tau\sin\theta_\tau,0,\beta_\tau\cos\theta_\tau),\nonumber \\ q  &=& \omega(1,\sin\theta_\gamma,0,\cos\theta_\gamma),\quad p_4=p_1+p_2-p_3-q.
\eea
We only use this parameterization to analyze the NLP result since the explicit $1/\omega$ factor at LP will have corrections at NLP. At the accuracy of our calculation, in $M_{\text{NLP}}$ we can set 
\bea
\label{eq:born}
E_3=E_4=\sqrt{s}/2, \qquad \vec{p}_3=-\vec{p}_4, \qquad \beta_\tau=\beta_{\bar{\tau}}=\beta=\sqrt{1-\frac{4m_\tau^2}{s}}\, .
\eea
In this approximation, the solution of Eq.~\eqref{eq:lprz} can be written as 
\bea
\label{eq:bornzero}
\cot(\theta_\gamma-\theta_\tau)=\frac{\beta\sin\theta_\tau\pm\sqrt{\beta^2\sin^2\theta_\tau
-4(1-\beta^2)(1-\beta\cos\theta_\tau)}}{2(1-\beta^2)}\,.
\eea
There are four radiation zeros for each value of $\theta_\tau$ when $\beta^2\sin^2\theta_\tau
-4(1-\beta^2)(1-\beta\cos\theta_\tau)>0$, as we show in Fig.~\ref{fg:2zeros}. The radiation zeroes exist for a wide range of $\tau$ kinematics. We show the radiation zero for both the Belle-II kinematics and for $\sqrt s=100$ GeV to illustrate the high-energy limit. In the high energy limit, the other two branches of the radiation zero appear when the photon is collinear to  $\tau$ or $\bar{\tau}$, and are therefore hard to measure. In the following we can also see that the NLP matrix element is proportional to $m_\tau/s$ and is therefore strongly suppressed in the high-energy limit at fixed non-collinear angles.

\begin{figure}[t]
    \centering
    \includegraphics[width=0.45\linewidth]{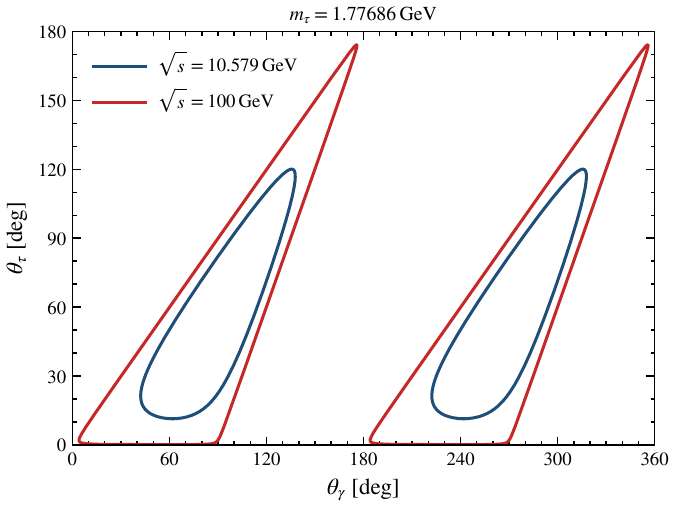}
    \caption{Positions of the leading soft radiation zeros in the $\theta_\tau,\theta_\gamma$ plane, obtained from Eq.~\eqref{eq:bornzero}. Results are shown for Belle-II kinematics and for $\sqrt s=100$ GeV to illustrate the high-energy limit.}
    \label{fg:nearzero}
\end{figure}

At the radiation zero we note that $M_{\text{NLP}}^{+-h_\lambda}\big|_0=M_{\text{NLP}}^{-+h_\lambda}\big|_0=0$ up to a $O(\omega/s)$ correction for both photon helicities. Here $\big|_0$ denotes evaluation at the radiation zero. We derive this result in Appendix~\ref{sec:derive}. 

For equal $\tau$ helicities $h=\pm1$, opposite electron and positron helicities $\eta=h_{e^-}=-h_{e^+}=\pm1$, and photon helicity $\lambda=\pm1$, its magnitude is
\bea
\left|M_{\mathrm{NLP},\eta,-\eta}^{hh\lambda}\right|\big|_0
=\frac{2\sqrt{2}m_\tau}{s \left[1-\beta^2\cos^2(\theta_\gamma-\theta_\tau)\right]}
(1-h\lambda\beta)(1+\eta\lambda\cos\theta_\gamma).
\eea
For a fixed photon helicity, when $\sqrt{s}\gg m_\tau$, the $h=\lambda$ amplitude is suppressed by $(1-\beta)/(1+\beta)$ relative to $h=-\lambda$, and the dominant components are $|++-\rangle$ and $|--+\rangle$. 

For the remainder of this section, we consider the $e^-_L e^+_R$ initial state without loss of generality for the pure-state entanglement analysis. The two dominant components share the same factor $1+\beta$, so that
\bea
\label{eq:rel}
\left|\frac{M_{\mathrm{NLP}}^{++-}\big|_0}{M_{\mathrm{NLP}}^{--+}\big|_0}\right|=\frac{1+\cos\theta_\gamma}{1-\cos\theta_\gamma}\,.
\eea

\begin{figure}[t]
    \centering
    \includegraphics[width=0.45\linewidth]{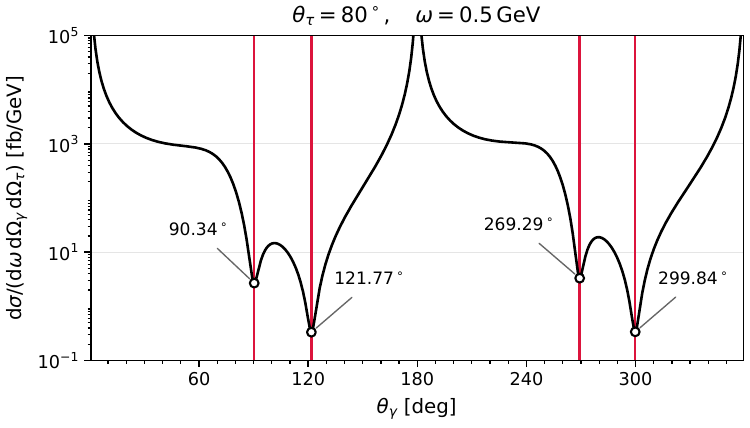}
    \includegraphics[width=0.45\linewidth]{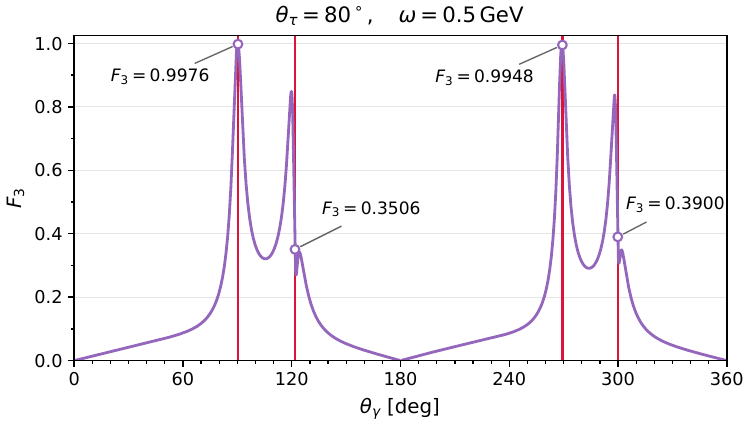}
    \caption{Scan of the cross section and $F_3$ at $\sqrt{s}=10.58\,\mathrm{GeV}$, $\theta_\tau=80^\circ$, and $\omega=0.5\,\mathrm{GeV}$. The red lines mark the locations of the radiation zeroes.}
    \label{fg:2zerossig}
\end{figure}

We can better see these features from Fig.~\ref{fg:2zerossig} and Fig.~\ref{fg:2zeros}. In Fig.~\ref{fg:2zerossig}, we scan in $\theta_\gamma$ with $\theta_\tau=80^\circ$, $\sqrt{s}=10.58\,\mathrm{GeV}$, $\omega=0.5\,\mathrm{GeV}$. In the left panel we show the scan of the cross section, while on the right we show the corresponding values of $F_3$. The four local minima corresponding to radiation zeroes are visible on the plot, considering the two with $\theta_\gamma<\pi$, one is at approximately $90.3^\circ$, the other is at approximately $122^\circ$. The red lines mark the two local minima of the full differential cross section. We will denote them with red lines as well in Fig.~\ref{fg:2zeros}. The location of radiation zero is consistent with the predicted in Fig.~\ref{fg:nearzero}. 

In the right panel, we note that the peaks of $F_3$ appear near the radiation zero. We can understand this by looking at the helicity-resolved matrix element square shown in Fig.~\ref{fg:2zeros}. As expected, at the radiation zero highlighted by the red line, the contribution is dominated by the $++-$ and $--+$ states. We can calculate the relative size between these two contributions from Eq.~\eqref{eq:rel}. For the radiation zero at $122^\circ$, we get $\left|{M_{\mathrm{NLP}}^{++-}\big|_0}\right|/\left|{M_{\mathrm{NLP}}^{--+}\big|_0}\right|=0.310 $. The state can then be approximately written as $0.296|++-\rangle+e^{i\phi}0.955|--+\rangle$ with an overall phase. The corresponding $F_3\sim0.32$, consistent with the lower panel of Fig.~\ref{fg:2zerossig}. $F_3$ reaches about $0.85$ about $2^\circ$ below the cross-section minimum. At this angle, the leading soft term is not negligible and brings the two amplitudes closer in magnitude.

Similarly for the radiation zero at $90.3^\circ$, we get 
$
\left|{M_{\mathrm{NLP}}^{++-}\big|_0}\right|/\left|{M_{\mathrm{NLP}}^{--+}\big|_0}\right|=0.988.
$
This state can then be approximately written as $(|++-\rangle+e^{i\phi}|--+\rangle)/\sqrt{2}$ with an overall phase. This is a GHZ state and $F_3\sim1$ as shown in the lower panel of Fig.~\ref{fg:2zerossig}. It is the dominance of these helicity amplitudes sub-leading in the high-energy limit, together with their angular dependence, that leads to the appearance of a GHZ-like state. The two additional radiation zeros in the region $\theta_\gamma>\pi$ can be analyzed in the same way, and we do not discuss them further here.

\section{Numerics}
\label{sec:num}

We study entanglement at the radiation zero assuming Belle-II kinematics. We set $\sqrt s=10.579$~GeV, $m_\tau=1.77686$~GeV, which leads to a $\tau$-lepton velocity of $\beta_\tau=0.9419$ in the CM frame. As pointed out in the previous section the enhancement of entanglement occurs from helicity amplitudes that are suppressed in the massless limit, making this effect most relevant for lower-energy colliders where the $\tau$-mass is a non-negligible effect. 

We begin by studying measures of bipartite and tripartite entanglement near and away from each of the two radiation zeroes found before. We set the polar angle of the $\tau$ to $\theta_\tau=30^\circ$ and set the initial-state polarizations to $\lambda_{e^-}=-1, \lambda_{e^+}=+1$ for illustrative purposes. We show in Table~\ref{tab:raz} the following measures of bi-partite entanglement: ${\mathcal C}_{\tau\bar\tau}$, ${\mathcal C}_{\tau\gamma}$ and ${\mathcal C}_{\bar\tau\gamma}$. We also show $F_3$ as a measure of tripartite entanglement, and the differential cross section. We choose several representative phase-space points near the two radiation zeroes, as well as points away from both zeroes.

\begin{table}[htbp]
\centering
\begin{tabular}{ccccccc}
\hline\hline
$\omega$ (GeV) & $\theta_\gamma$ & $\dfrac{d\sigma}{d\omega\, d\Omega_\gamma\, d\Omega_\tau}$ (fb/GeV)
 & ${\mathcal C}_{\tau\bar\tau}$ & ${\mathcal C}_{\tau\gamma}$ & ${\mathcal C}_{\bar\tau\gamma}$ & $F_3$ \\
\hline
\multicolumn{7}{l}{\textit{Away from the radiation zeros}} \\
0.05 & $150^\circ$   & $2.1\times10^{3}$ & 0.13 & 0.00 & 0.01 & 0.00 \\
0.5  & $150^\circ$   & $2.5\times10^{2}$ & 0.14 & 0.01 & 0.05 & 0.01 \\
\hline
\multicolumn{7}{l}{\textit{Radiation zero near $\theta_\gamma=90^\circ$}} \\
0.05 & $95.9^\circ$  & $0.0086$          & 0.06 & 0.01 & 0.01 & 0.96 \\
0.5  & $90.9^\circ$  & $0.11$            & 0.07 & 0.04 & 0.04 & 1.00 \\
\hline
\multicolumn{7}{l}{\textit{Radiation zero near the $\tau$}} \\
0.05 & $45.3^\circ$  & $0.30$            & 0.07 & 0.05 & 0.00 & 0.12 \\
0.5  & $49.4^\circ$  & $2.0$             & 0.06 & 0.12 & 0.01 & 0.16 \\
\hline\hline
\end{tabular}
\caption{Differential cross section and entanglement measures for $e^-_L e^+_R\to\tau\bar\tau\gamma$
at the Belle-II energy $\sqrt{s}=10.579$~GeV, with the $\tau$ at polar angle $\theta_\tau=30^\circ$ and the photon in the
beam--$\tau$ plane at polar angle $\theta_\gamma$, both measured from the $e^-$ beam in the
center-of-mass frame. The cross section is differential in the photon energy $\omega$, the photon
solid angle $\Omega_\gamma$ and the $\tau$ solid angle $\Omega_\tau$. The radiation zeros are the
local minima of the cross section with respect to the photon direction.}
\label{tab:raz}
\end{table}

We begin by discussing points away from the radiation zeroes. The $F_3$ entanglement measure is near zero, indicating a lack of tripartite entanglement. This picture changes significantly at the radiation zero near $\theta_\gamma=90^\circ$. The bipartite entanglement measures are near zero, and $F_3$ approaches its maximum value of one, indicating a GHZ-like state. While the differential cross section becomes small, it is non-zero and measurable given the integrated luminosity of Belle-II. We can also calculate the relative size between these two contributions from Eq.~\eqref{eq:rel}. For the radiation zero at $45.3^\circ$, we get $\left|{M_{\mathrm{NLP}}^{++-}\big|_0}\right|/\left|{M_{\mathrm{NLP}}^{--+}\big|_0}\right|\approx 5.74 $. The state can then be approximately written as $0.986|++-\rangle+e^{i\phi}0.171|--+\rangle$ with an overall phase. The corresponding $F_3\sim0.11$, consistent with the table.

\begin{figure}[htbp]
    \centering
    \includegraphics[width=0.7\linewidth]{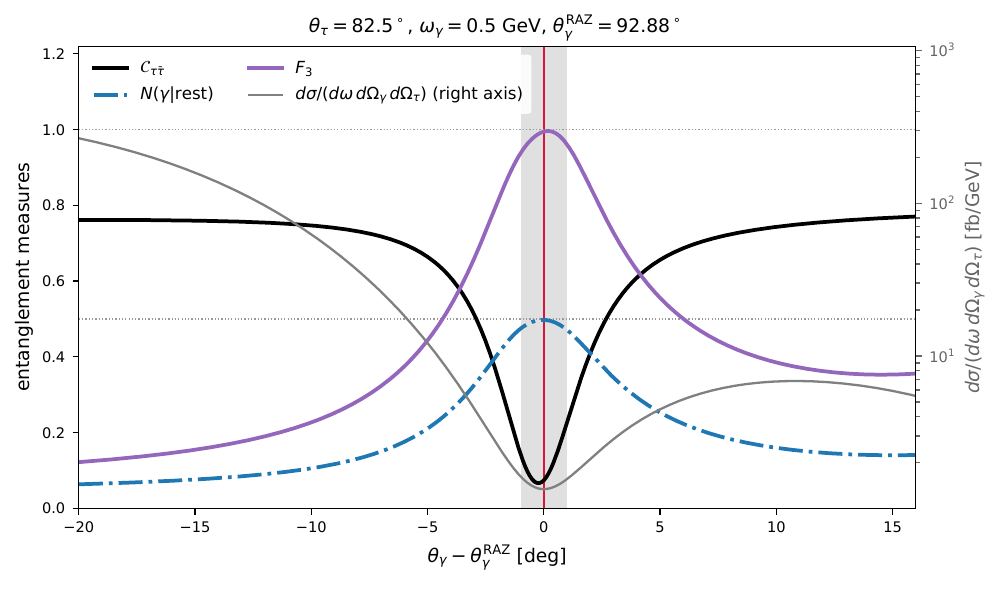}
    \caption{Entanglement measures and differential cross section at $\theta_{\tau}=82.5^\circ$ and $\omega_\gamma=0.5\,\mathrm{GeV}$.
    The black curve shows the $\tau\bar\tau$ concurrence $\mathcal C_{\tau\bar\tau}$, and the purple curve shows the tripartite entanglement measure $F_3$. The blue dash-dotted curve shows the negativity $N(\gamma|\mathrm{rest})$ between the photon and the $\tau\bar\tau$ pair.
    The gray curve shows the differential cross section on the right axis.}
    \label{entfig}
\end{figure}

\begin{figure}[t]
    \centering
    \includegraphics[width=0.9\linewidth]{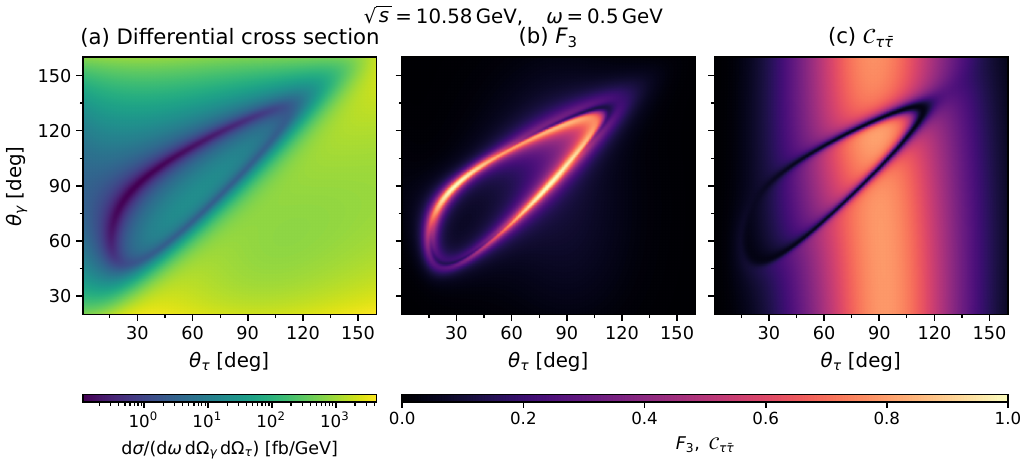}
    \caption{Angular dependence of (a) the differential cross section $\mathrm{d}\sigma/(\mathrm{d}\omega\,\mathrm{d}\Omega_\gamma\,\mathrm{d}\Omega_\tau)$, (b) $F_3$, and (c) the $\tau\bar{\tau}$ concurrence $\mathcal{C}_{\tau\bar{\tau}}$.}
    \label{fg:ghzfig}
\end{figure}

In Fig.~\ref{entfig} we show the entanglement measures and differential cross section at $\theta_{\tau}=82.5^\circ$ and $\omega_\gamma=0.5\,\mathrm{GeV}$. We can see $F_3$ approaches unity and the negativity $N(\gamma|\mathrm{rest})$ approaches its maximum value $1/2$. This means near the radiation zero we achieve a GHZ-like state. The $\tau\bar\tau$ concurrence $\mathcal C_{\tau\bar\tau}\sim0$ near the radiation zero. This suppression is consistent with the ideal GHZ state, whose two-qubit reduced states all have zero concurrence.

The left panel of Fig.~\ref{fg:ghzfig} shows the cross section as a function of the CM polar angle of the photon and the $\tau$-lepton. The shape and depth of the two radiation zeroes are clear from the plot, and the same shape is consistent with Fig.~\ref{fg:2zeros}. This  validates the soft photon approximation used to obtain the simple analytic solution before. In the middle panel of Fig.~\ref{fg:ghzfig} we show the tripartite entanglement measure $F_3$. We can see near the radiation zero,  tripartite entanglement is strongly enhanced. We can also see that around $\theta_\gamma=90^\circ$, the tripartite entanglement is close to one, indicating the final state is close to a GHZ-like state. In the right panel of Fig.~\ref{fg:ghzfig}, we can also see $C_{\tau\bar{\tau}}$ is suppressed at the radiation zero. We only show the branch with $\theta_\gamma<\pi$, the pattern for $\theta_\gamma>\pi$ is the same. 

\subsection{Observability of the effect}

Since we are studying $\tau$ production near radiation zeroes where the event rate approaches zero, one may worry that even with a large $F_3$ the effect is unobservable due to a vanishing event rate. We show here that there are good prospects for studying this effect given expected Belle-II integrated luminosities.
This is driven by the fact that the GHZ-like entanglement persists a measurable distance away from the radiation zeroes. To show this we begin by defining a cone of half-angle $\delta$ in the photon direction centered at each radiation zero, with the intent of averaging $F_3$ over this cone. For a given $\tau$ direction and photon energy, the tripartite entanglement is large in a region right around the zero. On each side of the beam axis in the beam-$\tau$ plane we select the zero leading to the largest $F_3$, and sum the results of both zeroes, so that we combine the results of the two largest out of the four existing zeroes. We focus on the $\tau$ region $75^\circ<\theta_\tau<90^\circ$ and $0.4<\omega<0.8$~GeV, where $F_3$ at the zero reaches or exceeds $0.9$. 
For unpolarized beams, we construct the spin density matrix at each phase-space point by summing the $e^-e^+$ helicity contributions. Both contributions are obtained from the full complex amplitudes in the same outgoing helicity basis. We evaluate the convex-roof $F_3$~\cite{Jin:2022kxb} of this mixed state at each point and report its cross-section-weighted average in Table~\ref{tab:rates}. For this process in the massless-electron limit the convex roof equals the helicity-averaged result for $F_3$.

\begin{table}[htbp]
\centering
\begin{tabular}{ccccc}
\hline\hline
$\delta$ & $\sigma$ (fb) & $\langle F_3\rangle$ & Produced & Both $\tau$ hadronic \\
\hline
$3^\circ$ & 0.041 & 0.76 & 2040  & 420  \\
$4^\circ$ & 0.103 & 0.65 & 5140  & 1050 \\
$5^\circ$ & 0.221 & 0.55 & 11050 & 2270 \\
\hline\hline
\end{tabular}
\caption{Cross section, average tripartite entanglement and event yields at $50~{\rm ab}^{-1}$
in cones of half-angle $\delta$ around the two radiation zeros with large $F_3$, for
$75^\circ<\theta_\tau<90^\circ$, $0.4<\omega<0.8$~GeV and unpolarized beams at
$\sqrt{s}=10.579$~GeV. ``Produced'' is the yield before any efficiencies; the last column
requires both $\tau$ leptons to decay to $\pi\nu$, $\pi\pi^0\nu$ or $\pi^-\pi^+\pi^-\nu$.}
\label{tab:rates}
\end{table}

The results are shown in Table~\ref{tab:rates} for an integrated luminosity of $50~{\rm ab}^{-1}$, consistent with Belle-II expectations, and for several choices of opening angle $\delta$. A smaller opening angle leads to a larger $\langle F_3\rangle$ but a smaller event rate, so there is a trade-off.  $\langle F_3\rangle$ decreases from $0.76$ at $\delta=3^\circ$ to $0.55$ at $\delta=5^\circ$, while the number of produced events increases from approximately 2040 to 11050. In addition to the raw event rate we also give the yield after both $\tau$ leptons to decay to one of the three hadronic channels. No other selection or reconstruction efficiencies are included. 

The measurement of $F_3$ requires determination of the photon polarization, and how to accomplish that is an important issue to discuss. One possibility is to measure the photon linear polarization from the azimuthal distribution of $e^+e^-$ pairs produced by conversion in the detector material~\cite{deLima:2026jbb}. The estimated conversion probability at Belle-II is about 2\% of the full event rate, which  restricts the effective statistics. A more promising possibility is to infer the photon's entanglement from the $\tau$ leptons,
similar to a proposal for radiative $B$ decays~\cite{Pakhlov:2021yme}. For
a pure three-qubit state the reduced density matrices of complementary subsystems have equal
purity, so that
$\mathcal{C}_{\gamma(\tau\bar\tau)}=\sqrt{2\left(1-{\rm Tr}\,\rho_{\tau\bar\tau}^2\right)}$,
and the remaining one-to-other concurrences follow from the single-$\tau$ spin states. $F_3$ is
then determined by $\tau$-pair spin tomography~\cite{Ehataht:2023zzt}. We leave a more detailed study of these approaches to future work.

\section{Conclusion}
\label{sec:con}

We have studied the tripartite entanglement structure of the spin state produced near radiation zeroes for $e^+e^-\rightarrow\tau\bar{\tau}\gamma$. Our result is based on the tree-level QED helicity amplitudes. For fixed initial helicities, we quantify genuine tripartite entanglement using the concurrence-triangle measure $F_3$. Around the radiation zero, the cancellation of the leading soft amplitude drives a sharp enhancement of tripartite entanglement. In the soft limit along the zero, the next-to-leading-power contribution also vanishes for opposite $\tau$ helicities, leaving mass-dependent amplitudes for equal $\tau$ helicities. In the Belle-II kinematics, because of the suppression of the relative factor $(1-\beta_\tau)/(1+\beta_\tau)$, the dominant components are $|++-\rangle$ and $|--+\rangle$. For the radiation zero near $\theta_\gamma=90^\circ$, their magnitudes become comparable, producing a GHZ state. For a left hand $e^-$, a right hand $e^+$, $\omega=0.5\,\mathrm{GeV}$, we find $F_3\simeq0.9985$ at a sampled point near one of the radiation zero with $\theta_\gamma\sim90^\circ$. We argued that this effect is measurable. At $50\,\mathrm{ab}^{-1}$, with unpolarized beam, the tree-level rates give several hundred to a few thousand events depending on the cone size. The mean tripartite entanglement $F_3$ remains sizable across these angular windows.

Our results identify a mechanism for producing a GHZ state in single-photon radiative processes. Near a radiation zero, the cancellations that suppress the differential rate can also simplify the helicity structure of the final state, leaving only a few dominant amplitudes. This simplification makes it easier to study QIS and suggests looking for GHZ states in other radiative processes where the helicity-conserving amplitude is zero, and the subleading helicity-flip amplitudes exposes this entangled structure.

\begin{acknowledgments}
H.~C. thanks Shanshan Hu and Yang-Li Zeng for their help in establishing Codex. H.~C. and F.~P. are supported by the U.S. Department of Energy, Office of High Energy Physics, under contract No.~DE-SC0010143, H.~C. is partially supported by a CFNS Joint Postdoctoral Fellowship. H.~C. and F.~P.  also acknowledge the support in part through the computational resources and staff contributions provided for the Quest high performance computing facility at Northwestern University which is jointly supported by the Office of the Provost, the Office for Research, and Northwestern University Information Technology. Claude and OpenAI Codex were used to assist with calculations and numerical code development; the code is independently tested and validated by the authors.
\end{acknowledgments}

\begin{widetext} 
\appendix 

\section{Next-to-leading power contribution for different $\tau\bar{\tau}$ helicity}\label{sec:derive}

We separate ${\cal S}_{\text{LP}}$ from Eq.~\eqref{eq:M_LP} into a initial and final part to get
\begin{align}
{\cal{S}}_{\text{LP,I}}^\mu
&\equiv-\frac{p_1^\mu}{p_1\cdot\hat{q}}+\frac{p_2^\mu}{p_2\cdot\hat{q}}\nn
\\
{\cal{S}}_{\text{LP,F}}^\mu
&\equiv\frac{ p_3^\mu}{p_3\cdot\hat{q}}-\frac{p_4^\mu}{p_4\cdot\hat{q}}
\end{align}
We also define the vector and axial kernels as 
\begin{align}
\label{eq:A1}
  A_e^\mu=\bar{v}_e\gamma^\mu\gamma^5 u_e,
  \quad A_\tau^\mu=\bar{u}_\tau\gamma^\mu\gamma^5 v_\tau
  ,\quad
    V_e^\mu=\bar{v}_e\gamma^\mu u_e,
  \quad V_\tau^\mu=\bar{u}_\tau\gamma^\mu v_\tau\,.
\end{align}
For massless electrons and assuming the CM frame, using the approximation in Eq.~\eqref{eq:born}, the following relation holds
\begin{align}
\label{eq:A2}
  A_e^0=V_e^0=V_e\cdot\hat{z}=V_\tau^0=0,\quad 
   A_e^\mu=i\epsilon^{0\mu\nu\alpha}\hat z_\nu  V_{e,\alpha}\,.
\end{align}
 When the helicities of the $\tau\bar{\tau}$ pair are different, we have
\begin{align}
\label{eq:A3}
  V_\tau\cdot\hat p_3=0,\quad 
  A_\tau^\mu=-i\epsilon^{0\mu\nu\alpha}{\hat{p}}_{3,\nu} {V}_{\tau,\alpha}
\end{align}
with $\hat{p}\equiv p/E$

In the following, we use the $\epsilon^*_0=0$ gauge and study the radiation zero point at Leading power. 
From ${\cal{S}}_{\text{LP}}\cdot\epsilon^*= 0$ we get
\begin{align}
\vec{\cal{S}}_{\text{LP}}\propto \vec{\hat{q}}\,.
\end{align}
Therefore $S_{\text{LP}}^0=\vec{\hat{q}}\cdot\vec{\cal{S}}_{\text{LP}}$,
where 
\begin{align}
  S_{\text{LP}}^0&=\frac{\sqrt{s}\omega}{2}\left(-\frac{1}{p_1\cdot q}+\frac{1}{p_2\cdot q}
  +\frac{1}{p_3\cdot q}-\frac{1}{p_4\cdot q}\right)\,,
\nn
\\
  \mathcal{S}_{\text{LP}}^i&=S_{\text{LP}}^0\hat{q}^i\,.
\end{align}
We use the equation
\begin{align}
 \gamma^\mu
 \slashed{q} \slashed{\varepsilon}^{*}(q)&=
 q^\mu\slashed{\varepsilon}^{*}(q)-\varepsilon^{*\mu}(q)\slashed{q}
 +i\varepsilon^{\mu\nu\alpha\beta}q_\nu\varepsilon^*_\alpha(q)\gamma_\beta\gamma_5\, ,\nn
\\
\slashed{\varepsilon}^{*}(q)
 \slashed{q}
 \gamma^\mu&=q^\mu\slashed{\varepsilon}^{*}(q)
 - \varepsilon^{*\mu}(q)\slashed{q}
 -i\varepsilon^{\mu\nu\alpha\beta}q_\nu\varepsilon^*_\alpha(q)\gamma_\beta\gamma_5\,.
\end{align}
We can derive the next-to-leading power contribution without $\gamma_5$ in Eq.~\eqref{eq:M_LP}; naming it $M_{\text{NLP,V}}$ we have
\begin{align}
  M_{\text{NLP,V}}
  =\frac{1}{s\sqrt{s}}(\mathcal{S}_{\text{LP,F},\mu}\varepsilon^{*}_\nu(q)-\mathcal{S}_{LP,I,\nu}\varepsilon^{*}_\mu(q))V_{e}^\mu V_{\tau}^\nu+\frac{2}{\sqrt{s}}M_0\left(-\frac{p_1^\nu}{p_1\cdot \hat{q}}+\frac{p_2^\nu}{p_2\cdot \hat{q}}\right)\varepsilon^*_\nu(q)\,.
\end{align}
Using Eq.~\eqref{eq:A1}-\eqref{eq:A3}, we get
\begin{align}
  &iV_{e,\mu}\epsilon^{\mu\nu\alpha\beta}q_\nu\varepsilon^*_\alpha
  A_{\tau,\beta}=-\omega (V_{e,\mu}{V}_{\tau}^\mu\varepsilon^*_\alpha {\hat{p}}_{3}^\alpha
  -V_{e,\mu} {\hat{p}}_{3}^\mu \varepsilon^*_\alpha{V}_{\tau}^\alpha)\,,\nn
  \\
  &iV_{\tau,\mu}\epsilon^{\mu\nu\alpha\beta}q_\nu\varepsilon^*_\alpha
  A_{e,\beta} =\omega (V_{\tau,\mu}{V}_{e}^\mu\varepsilon^*_\alpha {\hat{z}}^\alpha
  -V_{\tau,\mu} {\hat{z}}^\mu \varepsilon^*_\alpha{V}_{e}^\alpha)\,.
\end{align}
The axial contribution to the next-to-leading power matrix element $M_{\text{NLP}}$ in Eq.~\eqref{eq:M_LP} will have the following form
\begin{align}
  M_{\text{NLP,A}}=-\frac{1}{ \sqrt{s}}M_0
  \varepsilon^{*}_\nu
  \left(-{\cal{S}}_{\text{LP,F}}^\nu+{\cal{S}}_{\text{LP,I}}^\nu\right)
  -\frac{1}{s\sqrt{s}}V_{e}^\mu V_{\tau}^\nu
  \left(
  \varepsilon^*_\nu{\cal{S}}_{\text{LP,F},\mu}
-{\varepsilon^*_\mu}{\cal{S}}_{\text{LP,I},\nu}
  \right)\,.
\end{align}
Using ${\cal{S}}_{\text{LP,F}}^\nu\varepsilon^*_\nu=-{\cal{S}}_{\text{LP,I}}^\nu\varepsilon^*_\nu$ at radiation zero
and putting everything together, we have
\begin{align}
M_{\text{NLP}}=&\frac{1}{s\sqrt{s}}(\mathcal{S}_{\text{LP,F},\mu}\varepsilon^{*}_\nu-\mathcal{S}_{\text{LP,I},\nu}\varepsilon^{*}_\mu)V_{e}^\mu V_{\tau}^\nu
  -\frac{1}{s\sqrt{s}}V_{e}^\mu V_{\tau}^\nu
  \left(
  \varepsilon^*_\nu{\cal{S}}_{\text{LP,F},\mu}
-{\varepsilon^*_\mu}{\cal{S}}_{\text{LP,I},\nu}
  \right) \nn
  \\
  -&\frac{2}{ \sqrt{s}}M_0
  \varepsilon^{*}_\nu
  {\cal{S}}_{\text{LP,I}}^\nu
  +\frac{2}{\sqrt{s}}M_0{\cal{S}}_{\text{LP,I}}^\nu\varepsilon^*_\nu\nn
  \\
  =&0
\end{align}
\end{widetext} 
\bibliographystyle{h-physrev}   
\bibliography{refs}
\end{document}